\documentclass{article}
\usepackage{amsmath,graphicx,mlspconf}

\usepackage{bm}
\usepackage{txfonts}
\usepackage{relsize}
\usepackage{udline}
\usepackage[linesnumbered,lined,ruled]{algorithm2e}
\setulminsep{1.2ex}{0.6ex}
\setulsep{-2.5pt}

\newcommand{\nr}{\mathbb{R}_{\scalebox{0.6}{$\geq0$}}}
\newcommand{\nc}{\mathbb{C}}

\makeatletter
 \def\Hline{%
 \noalign{\ifnum0=`}\fi\hrule \@height 1.4pt \futurelet
 \reserved@a\@xhline}
\makeatother

\AtBeginDvi{}

\copyrightnotice{U.S.\ Government work not protected by U.S.\ copyright}

\copyrightnotice{978-1-5090-6341-3/17/\$31.00 {\copyright}2017 Crown}

\copyrightnotice{978-1-5090-6341-3/17/\$31.00 {\copyright}2017 European Union}

\copyrightnotice{Preprint manuscript of IEEE MLSP 2017 {\copyright}2017 IEEE}

\toappear{Preprint manuscript of 2017 IEEE International Workshop on Machine Learning for Signal Processing}

\title{Independent Low-Rank Matrix Analysis Based on Complex Student's $t$-Distribution for Blind Audio Source Separation}
\name{{\em Shinichi Mogami}$^{\ \! 1}$,
          {\em Daichi Kitamura}$^{\ \! 1}$,
          {\em Yoshiki Mitsui}$^{\ \! 1}$,
          {\em Norihiro Takamune}$^{\ \! 1}$, \\
          {\em Hiroshi Saruwatari}$^{\ \! 1}$, 
		 {\em Nobutaka Ono}$^{\ \! 2}$}
\address{
    $^1$ The University of Tokyo, Tokyo, Japan\\
    $^2$ National Institute of Informatics, Tokyo, Japan}

\begin{document}
\ninept

\maketitle
\begin{abstract}
In this paper, we generalize a source generative model in a state-of-the-art blind source separation (BSS), independent low-rank matrix analysis (ILRMA). ILRMA is a unified method of frequency-domain independent component analysis and nonnegative matrix factorization and can provide better performance for audio BSS tasks. To further improve the performance and stability of the separation, we introduce an isotropic complex Student's $t$-distribution as a source generative model, which includes the isotropic complex Gaussian distribution used in conventional ILRMA. Experiments are conducted using both music and speech BSS tasks, and the results show the validity of the proposed method.
\end{abstract}
%

\section{Introduction}
\label{sec:intro}
Blind source separation (BSS) is a technique for extracting specific sources from an observed multichannel mixture signal 
without knowing a priori information about the mixing system. The most popular algorithm for BSS is called 
independent component analysis (ICA)~\cite{PComon1994_ICA}, which assumes statistical independence between the sources 
and estimates the demixing system. 
In particular, BSS for audio signals has been well studied. For a mixture of audio signals, since the sources are convolved 
owing to the room reverberation, ICA is often applied to the time-frequency domain signal, which is called the spectrogram 
obtained by a short-time Fourier transform (STFT). 
Frequency-domain ICA (FDICA)~\cite{PSmaragdis1998_BSS,HSaruwatari2006_FDICA} 
independently applies ICA to the time-series signals in each frequency, then the permutation of the estimated 
signals is aligned on the basis of several criteria. As an elegant solution of this permutation alignment problem, independent vector 
analysis (IVA)~\cite{TKim2007_IVA} was proposed, which assumes higher-order dependences among the frequency components in 
each source, thus avoiding the permutation problem. 
In~\cite{NOno2011_AuxIVA}, fast and stable optimization of IVA (AuxIVA) was derived using an auxiliary function technique that is 
also known as a majorization-minimization (MM) algorithm~\cite{DRHunter2000_MMalgorithm}. 

As another means of audio source separation, nonnegative matrix factorization (NMF)~\cite{DDLee1999_NMF} has been a 
very popular approach during the last decade. NMF is a parts-based decomposition (low-rank 
approximation) of a nonnegative data matrix, which is typically a power or amplitude spectrogram, and the significant parts 
(bases and activations) can be used for source separation. 
Also, NMF can be statistically interpreted as a parameter estimation based on a generative model of 
data, and the distribution of the model defines a cost function (divergence) in NMF. For example, it was revealed that 
NMF based on Itakura--Saito divergence (ISNMF) assumes an isotropic complex Gaussian distribution independently defined in each 
time-frequency slot~\cite{CFevotte2009_ISNMF}. 
Recently, a new NMF based on an isotropic complex Cauchy distribution (Cauchy NMF)~\cite{ALiutkus2015_CauchyNMF} and its generalization, 
NMF based on a complex Student's $t$-distribution ($t$-NMF)~\cite{KYoshii2016_tNMF}, have been proposed. 
$t$-NMF includes both ISNMF and Cauchy NMF as special cases, 
and it has been reported that $t$-NMF provides better and more stable source separation for simple audio signals~\cite{KYoshii2016_tNMF}. 

For multichannel audio source separation, NMF has been extended to multichannel NMF 
(MNMF)~\cite{AOzerov2010_MNMF,HKameoka2010_MNMF,HSawada2013_MNMF}. MNMF employs a sourcewise spatial parameter, 
spatial covariance, that approximates the mixing system to achieve source separation. However, the separation performance 
of MNMF strongly depends on the initialization of the parameters because of the difficulty of the optimization. This problem was 
addressed by exploiting a complex Student's $t$-distribution as a source generative model in MNMF ($t$-MNMF)~\cite{KKitamura2016_tMNMF}, 
which may lead to initialization-robust optimization. 

\begin{figure}[t!]
\begin{center}
\vspace{2pt}
\includegraphics[width=7.2cm]{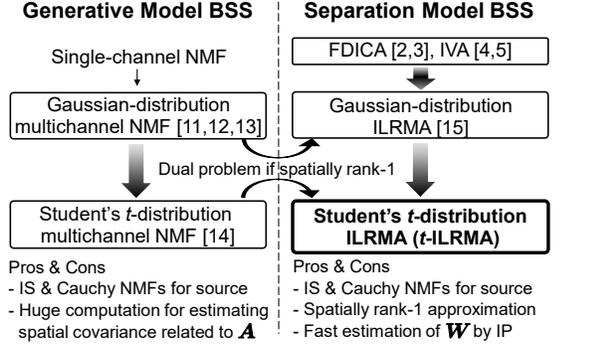}
\vspace{-8pt}
\caption{Relation among conventional methods and proposed ILRMA.}
\label{relationship}
\end{center}
\vspace{-18pt}
\end{figure}

NMF has been unified with the conventional ICA- or IVA-based techniques, which allows us to simultaneously model the sourcewise 
time-frequency structure and the statistical independence between sources. This state-of-the-art BSS is called {\it independent 
low-rank matrix analysis (ILRMA)}~\cite{DKitamura2016_ILRMA,YMitsui2017_sparseILRMA}, which is a natural extension of IVA 
from a vector to a low-rank matrix source model. ILRMA is equivalent to a special case of MNMF; ILRMA assumes that 
the mixing system is invertible and estimates the demixing system similarly to FDICA or IVA, whereas MNMF estimates the mixing 
system (spatial covariance) required for separation. For the optimization problem, ILRMA is much faster and more stable than MNMF. 
In this paper, we generalize the source generative model in ILRMA from 
the complex Gaussian distribution to the complex Student's $t$-distribution, which is expected to further improve the 
performance and stability of the parameter initialization. 
The relationship among the conventional methods and the proposed ILRMA is depicted in Fig.~\ref{relationship}.
As shown in this figure, the proposed ILRMA can be referred to as a new extension of conventional ILRMA as well as a 
computationally efficient solution for the dual problem of $t$-MNMF under a spatially rank-1 condition.

\section{Conventional methods}
\label{conv}
\subsection{Formulation}
\label{conv:formulation}
Let $N$ and $M$ be the numbers of sources and channels, respectively. The complex-valued source, observed, and estimated signals 
are defined as $\bm{s}_{ij} \!=\! (s_{ij,1}, \cdots, s_{ij,N})^{\rm T}$, $\bm{x}_{ij} \!=\! (x_{ij,1}, \cdots, x_{ij,M})^{\rm T}$, and 
$\bm{y}_{ij} \!=\! (y_{ij,1}, \cdots, y_{ij,N})^{\rm T}$, where $i\!=\!1, \cdots, I$; $j\!=\!1, \cdots, J$; $n\!=\!1, \cdots, N$; and 
$m\!=\!1, \cdots, M$ are the integral indexes of the frequency bins, time frames, sources, and channels, respectively, and $^{\rm T}$ 
denotes a transpose. We also denote the spectrograms of the source, observed, and estimated signals as 
$\bm{S}_{n}\ {\in \nc^{\scalebox{0.6}{$I{\times}J$}}}$, $\bm{X}_{m}\ {\in \nc^{\scalebox{0.6}{$I{\times}J$}}}$, and 
$\bm{Y}_{n}\ {\in \nc^{\scalebox{0.6}{$I{\times}J$}}}$, whose elements are $s_{ij,n}$, $x_{ij,m}$, and $y_{ij,n}$, respectively. In FDICA, IVA, 
and ILRMA, the following mixing system is assumed: 
\begin{align}
\bm{x}_{ij} = \bm{A}_{i}\bm{s}_{ij}, \label{mixture}
\end{align}
where $\bm{A}_{i}\!=\!(\bm{a}_{i,1}~\cdots~\bm{a}_{i,N})\ {\in \nc^{\scalebox{0.6}{$M{\times}N$}}}$ is a frequency-wise mixing matrix and 
$\bm{a}_{i,n}$ is the steering vector for the $n$th source. The assumption of the mixing system (\ref{mixture}) corresponds to restricting 
the spatial covariance in MNMF to a rank-1 matrix~\cite{DKitamura2016_ILRMA}. The estimated signal $\bm{y}_{ij}$ can be obtained 
by assuming $M\!=\!N$ and estimating the frequency-wise demixing matrix 
$\bm{W}_{i}\!=\!(\bm{w}_{i,1}~\cdots~\bm{w}_{i,N})^{\rm H}\!=\!\bm{A}_{i}^{-1}$ as 
\begin{align}
\bm{y}_{ij} = \bm{W}_{i}\bm{x}_{ij}, \label{separation}
\end{align}
where $\bm{w}_{i,n}$ is the demixing filter for the $n$th source and $^{\rm H}$ denotes a Hermitian transpose. FDICA, IVA, and ILRMA 
estimate both $\bm{W}_{i}$ and $\bm{y}_{ij}$ from only the observation $\bm{x}_{ij}$ assuming statistical independence between 
$s_{ij,n}$ and $s_{ij,n'}$, where $n'\!\neq\!n$.


\subsection{ILRMA}
\label{conv:ilrma}
ILRMA assumes the following time-varying distribution as the generative model of each source:
\begin{align}
\prod_{i,j} p(y_{ij,n}) =&\ \prod_{i,j} \frac{1}{\pi r_{ij,n}} \exp \left( -\frac{ |y_{ij,n}|^2 }{ r_{ij,n} } \right), \label{localGauss} \\ 
r_{ij,n} =&\ \sum_{l} t_{il,n}v_{lj,n}, 
\end{align}
where the local distribution $p(y_{ij,n})$ is defined as a circularly symmetric (isotropic) complex Gaussian distribution, i.e., the 
probability of $p(y_{ij,n})$ only depends on the power of the complex value $y_{ij,n}$. Also, $r_{ij,n}$ is a time-frequency-varying 
nonnegative variance and corresponds to the expectation of the power of $y_{ij,n}$, i.e., $r_{ij,n}\!=\!{\rm E}[|y_{ij,n}|^2]$. 
This is because $p(y_{ij,n})$ is isotropic in the complex plane. Moreover, $t_{il,n}$ and $v_{lj,n}$ are the NMF parameters called 
basis and activation, respectively, $l \!=\! 1, \cdots, L$ is the integral index, and $L$ is set to a much smaller value than 
$\min{(I,J)}$, which leads to the low-rank approximation. 
Since the variance $r_{ij,n}$ can fluctuate depending on the time frame, (\ref{localGauss}) becomes a non-Gaussian distribution. 
The negative log-likelihood function ${\cal L}$ based on (\ref{localGauss}) can be obtained as follows by assuming 
independence between each source and each time frame: 
\begin{align}
{\cal L} = {\rm const.} - 2J \sum_{i} \log |\det \bm{W}_{i}| + \sum_{i,j,n} \left( \log{ r_{ij,n} } + \frac{ |y_{ij,n}|^2 }{ r_{ij,n} } \right). \label{cost}
\end{align}
Regarding the estimation of $t_{il,n}$ and $v_{lj,n}$, the minimization of (\ref{cost}) is equivalent to the optimization in ISNMF that 
minimizes the Itakura--Saito divergence between $|\bm{Y}_{n}|^{.2}$ and $\bm{T}_{n}\bm{V}_{n}$, where $\bm{T}_{n}\ {\in \nr^{\scalebox{0.6}{$I{\times}L$}}}$ 
and $\bm{V}_{n}\ {\in \nr^{\scalebox{0.6}{$L{\times}J$}}}$ are the basis and activation matrices whose elements are $t_{il,n}$ and 
$v_{lj,n}$, and the absolute value and the dotted exponent for a matrix denote an element-wise absolute value and exponent, 
respectively.

\begin{figure}[t!]
\begin{center}
\vspace{-0pt}
\includegraphics[width=8.4cm]{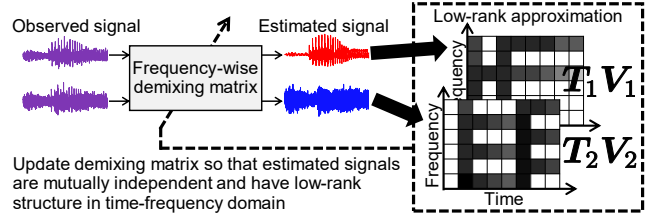}
\vspace{-8pt}
\caption{Conceptual model of ILRMA.}
\label{ILRMA}
\end{center}
\vspace{-24pt}
\end{figure}

Fig.~\ref{ILRMA} shows the conceptual model of ILRMA. When the original sources have a low-rank spectrogram $|\bm{S}_{n}|^{.2}$, 
the spectrogram of their mixture, $|\bm{X}_{m}|^{.2}$, should be more complicated, where the rank of $|\bm{X}_{m}|^{.2}$ will be 
greater than that of $|\bm{S}_{n}|^{.2}$. On the basis of this assumption, in ILRMA, the low-rank constraint for each estimated 
spectrogram $|\bm{Y}_{n}|^{.2}$ is introduced by employing NMF. The demixing matrix $\bm{W}_{i}$ is estimated so that the 
spectrogram of the estimated signal $|\bm{Y}_{n}|^{.2}$ becomes a low-rank matrix modeled by $\bm{T}_{n}\bm{V}_{n}$, whose rank 
is at most $L$. The estimation of $\bm{W}_{i}$, $\bm{T}_{n}$, and $\bm{V}_{n}$ can consistently be carried out by minimizing 
(\ref{cost}) in a fully blind manner. Note that ILRMA is theoretically equivalent to conventional MNMF only when the rank-1 
spatial model is assumed, which yields a stable and computationally efficient algorithm for ILRMA. This issue and the 
convergence-guaranteed fast update rules for $\bm{W}_{i}$, $\bm{T}_{n}$, and $\bm{V}_{n}$ can be found in~\cite{DKitamura2016_ILRMA}. 

\subsection{NMF and MNMF based on complex Student's $t$-distribution}
\label{conv:tNMF}

As revealed in~\cite{CFevotte2009_ISNMF}, ISNMF justifies the additivity of power spectra in the expectation sense using 
the stable property of a complex Gaussian distribution. Regarding the amplitude spectrogram, Cauchy 
NMF~\cite{ALiutkus2015_CauchyNMF} can be considered as a counterpart of ISNMF; the additivity of amplitude spectra is 
justified using the stable property of a complex Cauchy distribution. In~\cite{KYoshii2016_tNMF}, these theoretically justified 
NMFs were generalized by employing a complex Student's $t$-distribution, which includes the complex Gaussian and complex 
Cauchy distributions as special cases when the degree-of-freedom parameter $\nu\!>\!0$ is set to $\nu\!\rightarrow\!\infty$ 
and $\nu\!=\!1$, respectively. Although complex 
Student $t$-distributions with other values of $\nu$ do not have the stable property, $t$-NMF provides better and 
more robust source separation for simple audio signals when $\nu$ is approximately two. Also, the 
generalization of MNMF with a complex Student's 
$t$-distribution was proposed~\cite{KKitamura2016_tMNMF} with the aim of improving the robustness of the parameter 
initialization.

\section{Proposed method}
\label{prop}
\subsection{ILRMA based on complex Student's $t$-distribution}
\label{prop:tILRMA}

Motivated by the improvements in $t$-NMF, we propose the introduction of a complex Student's $t$-distribution as a source 
generative model in ILRMA ($t$-ILRMA), which is a generalization of conventional Gaussian ILRMA based on (\ref{localGauss}). 
The generative model in $t$-ILRMA is as follows: 
\begin{align}
\prod_{i,j} p(y_{ij,n}) =&\ \prod_{i,j} \frac{ 1 }{ \pi \sigma_{ij,n}^2 }\left( 1+\frac{2}{\nu} \frac{|y_{ijn}|^2}{\sigma_{ij,n}^2} \right)^{- \frac{2+\nu}{2}}, \label{localt} \\ 
\sigma_{ij,n}^{p} =&\ \sum_{l} t_{il,n}v_{lj,n}, \label{modelsigma}
\end{align}
where the local distribution $p(y_{ij,n})$ is defined as an isotropic complex Student's $t$-distribution, $\sigma_{ij,n}$ is a 
time-frequency-varying nonnegative scale and corresponds to an amplitude spectrum $|y_{ij,n}|$, and $p$ is a parameter 
that defines the domain of the NMF model $\bm{T}_{n}\bm{V}_{n}$ and should satisfy $1\!\leq\!p\!\leq\!2$. When $\nu\!\rightarrow\!\infty$ and $p\!=\!2$, (\ref{localt}) corresponds 
to the generative model in ISNMF, and when $\nu\!=\!1$ and $p\!=\!1$, (\ref{localt}) corresponds to the generative model in Cauchy NMF. 
The negative log-likelihood function based on (\ref{localt}) can be obtained as follows by assuming independence between each source 
and each time frame: 
\begin{align}
\nonumber {\cal L}_{t} =&\ {\rm const.} - 2J \sum_{i} \log |\det \bm{W}_{i}| \\
&\ + \sum_{i,j,n} \left[ \left( 1+\frac{\nu}{2} \right) \log{ \left( 1 + \frac{2}{\nu} \frac{|y_{ij,n}|^2}{ \sigma_{ij,n}^{2}} \right) } + 2\log \sigma_{ij,n}  \right]. \label{costt}
\end{align}
When $\nu\!\rightarrow\!\infty$ and $p\!=\!2$, (\ref{costt}) coincides with (\ref{cost}).

\subsection{Derivation of update rules for demixing matrix}
\label{prop:derivationW}

Similar to the derivation described in~\cite{DKitamura2016_ILRMA}, we apply an MM algorithm and iterative 
projection (IP)~\cite{NOno2011_AuxIVA} to derive the update rules for the demixing matrix $\bm{W}_{i}$ 
with a full guarantee of the monotonic convergence. 
IP was the method originally used to solve the simultaneous vector equations in AuxIVA, which are equivalent 
to the HEAD problem~\cite{AYeredor2009_HEADproblem}. 
Unlike the conventional MNMF methods such as that in~\cite{KKitamura2016_tMNMF} that estimate the mixing model 
$\bm{A}_{i}$ (not the demixing matrix $\bm{W}_{i}$), IP can lead to much faster and more stable estimation of $\bm{W}_{i}$ in 
BSS, as reported in~\cite{DKitamura2016_ILRMA,NOno2011_AuxIVA}. However, the major drawback of IP is the limited number 
of applicable functions; i.e., generally the term $|y_{ij,n}|^2 \!=\! |\bm{w}_{i,n}^{\rm H}\bm{x}_{ij}|^2$ should appear as is in the 
objective function, e.g., in (\ref{cost}) (should not appear as a part of variable inside a nonlinear function). 

For the $t$-ILRMA's cost function (\ref{costt}), whose $|y_{ij,n}|^2$ term is intrinsic, as a trick to enable the introduction of 
IP, we apply a tangent line inequality to the logarithm terms in (\ref{costt}). The tangent line inequality can be represented as 
\begin{align}
\log\left( \sum_{q} z_{q}\right) \leq \frac{1}{\lambda} \left( \sum_{q}z_{q} - \lambda \right) +\log \lambda, \label{tangent}
\end{align}
where $z_{q}$ is the original variable and $\lambda\!>\!0$ is an auxiliary variable. The equality of (\ref{tangent}) holds 
if and only if $\lambda \!=\! \sum_{q}z_{q}$. By applying (\ref{tangent}) to the second and third logarithm terms in (\ref{costt}), the following majorization function can be designed: 
\begin{align}
\nonumber {\cal L}_{t} \leq &\ {\rm const.} - 2J \sum_{i} \log |\det \bm{W}_{i}| \\
\nonumber &\ + \sum_{i,j,n} \left[ \left( 1+\frac{\nu}{2} \right) \frac{1}{\alpha_{ij,n}} \left( 1 + \frac{2}{\nu} \frac{|y_{ij,n}|^2}{\sigma_{ij,n}^{2}} -\alpha_{ij,n} \right) \right. \\
\nonumber &\ + \left( 1+\frac{\nu}{2} \right) \log \alpha_{ij,n} + \frac{2}{p\beta_{ij,n}} \left( \sum_{l} t_{il,n}v_{lj,n} -\beta_{ij,n}\right) \\
\nonumber &\ + \left. \frac{2}{p}\log \beta_{ij,n} \right] \\
\equiv&\ {\cal L}_{t}^{+}, \label{costtM}
\end{align}
where $\sigma_{ij,n}\!=\!(\sum_{l}t_{il,n}v_{lj,n})^{1/p}$ is partly substituted, $\alpha_{ij,n}, \beta_{ij,n}\!>\!0$ are auxiliary variables, and ${\cal L}_{t}$ and ${\cal L}_{t}^{+}$ become equal only when 
\begin{align}
\alpha_{ij,n} =&\ 1 + \frac{2}{\nu} \frac{|y_{ij,n}|^2 }{ \sigma_{ij,n}^{2}}, \label{alphaeq} \\
\beta_{ij,n} =&\ \sum_{l} t_{il,n}v_{lj,n}. \label{betaeq}
\end{align}
Because $|y_{ij,n}|^{2}\!=\!|\bm{w}_{i,n}^{\rm H}\bm{x}_{ij}|^{2}$ in (\ref{costtM}) exists outside the logarithm function, we can apply IP in analogy 
with the derivation in conventional ILRMA using (\ref{cost}). The majorization function (\ref{costtM}) can be reformulated as 
\begin{align}
\nonumber {\cal L}_{t}^{+} =&\ {\rm const.} - 2J \sum_{i} \log |\det \bm{W}_{i}| + J\sum_{i,n} \bm{w}_{i,n}^{\rm H} \bm{U}_{i,n} \bm{w}_{i,n} \\
\nonumber &\  + \sum_{i,j,n} \left[ \left( 1+\frac{\nu}{2} \right) \left( \alpha_{ij,n}^{-1} - 1 + \log \alpha_{ij,n} \right) \right. \\
&\ \left. + \frac{2}{p\beta_{ij,n}} \left(\sum_{l} t_{il,n}v_{lj,n} -\beta_{ij,n}\right ) + \frac{2}{p}\log \beta_{ij,n} \right], \label{costtMU} \\
\bm{U}_{i,n} =&\ \frac{1}{J}\left( \frac{2}{\nu} + 1 \right) \sum_{j} \frac{1}{\alpha_{ij,n}\sigma_{ij,n}^{2}}\bm{x}_{ij}\bm{x}_{ij}^{\rm H}. 
\end{align}
Since the majorization function (\ref{costtMU}) is the same form as that of AuxIVA with respect to $\bm{w}_{i,n}$, the following simultaneous equations are obtained: 
\begin{align}
\bm{w}_{i,k}^{\rm H} \bm{U}_{i,n} \bm{w}_{i,n} = \delta_{kn}, \label{wUw}
\end{align}
where $\delta_{kn}\!=\!1$ when $k\!=\!n$ and $\delta_{kn}\!=\!0$ when $k\!\neq\!n$. By applying IP to (\ref{wUw}), 
we can obtain the update rules for the demixing matrix as 
\begin{align}
\bm{w}_{i,n} \leftarrow&\ \left( \bm{W}_{i} \bm{U}_{i,n}\right)^{-1} \bm{e}_{n}, \label{updateW1} \\
\bm{w}_{i,n} \leftarrow&\ \frac{ \bm{w}_{i,n} }{ \sqrt{ \bm{w}_{i,n}^{\rm H}\bm{U}_{i,n}\bm{w}_{i,n} } }, \label{updateW2}
\end{align}
where $\bm{e}_{n}$ denotes the unit vector with the $n$th element equal to unity. After the update of $\bm{W}_{i}$, 
the separated signal $\bm{y}_{ij}$ should be updated as $y_{ij,n} \!\leftarrow\! \bm{w}_{ij,n}^{\rm H}\bm{x}_{ij}$.

\subsection{Derivation of update rules for NMF parameters}
\label{prop:derivationNMF}

The update rules for $t_{il,n}$ and $v_{lj,n}$ can be derived by the MM algorithm, which is a popular approach for NMF. 
To obtain the differentiable majorization function for NMF parameters in (\ref{costtM}), we apply Jensen's 
inequality to $\sigma_{ij,n}^{-2} = (\sum_{l} t_{il,n}v_{lj,n})^{-2/p}$. Jensen's inequality can be represented as 
\begin{align}
\left( \sum_{q} z_{q}\right)^{-2/p} = \left( \sum_{q}\mu_{q}\frac{z_{q}}{\mu_{q}} \right)^{-2/p} 
\leq \sum_{q} \mu_{q} \left( \frac{z_{q}}{\mu_{q}} \right)^{-2/p} 
= \sum_{q} \mu_{q}^{\frac{2}{p}+1} z_{q}^{-\frac{2}{p}}, \label{Jensen}
\end{align}
where $\mu_{q}\!>\!0$ is an auxiliary variable that satisfies $\sum_{q}\mu_{q}\!=\!1$. 
Note that the left-hand side of (\ref{Jensen}) is a convex function for the variable $z_{q}$ because we consider $1\!\leq\!p\!\leq\!2$. 
The equality of (\ref{Jensen}) holds if and only if $\mu_{q} \!=\! z_{q}/\sum_{q'}z_{q'}$.
By applying (\ref{Jensen}) to $\sigma_{ij,n}^{-2}\!=\!(\sum_{l}t_{il,n}v_{lj,n})^{-2/p}$ in (\ref{costt}), 
the following majorization function can be designed: 
\begin{align}
\nonumber {\cal L}_{t}^{+} \leq &\ {\rm const.} - 2J \sum_{i} \log |\det \bm{W}_{i}| \\
\nonumber & + \!\sum_{i,j,n} \left[ \left( 1+\frac{\nu}{2} \right) \frac{1}{\alpha_{ij,n}} \left( 1 + \frac{2}{\nu} |y_{ij,n}|^2 \!\sum_{l} \gamma_{ij,nl}^{\frac{2}{p}+1} t_{il,n}^{-\frac{2}{p}}v_{lj,n}^{-\frac{2}{p}} -\alpha_{ij,n} \right) \right. \\
\nonumber & + \left( 1+\frac{\nu}{2} \right) \log \alpha_{ij,n} + \frac{2}{p\beta_{ij,n}} \left( \sum_{l} t_{il,n}v_{lj,n} -\beta_{ij,n}\right) \\
\nonumber & + \left. \frac{2}{p}\log \beta_{ij,n} \right] \\
\equiv&\ {\cal L}_{t}^{++}, \label{costtMM}
\end{align}
where $\gamma_{ij,nl}\!>\!0$ is an auxiliary variable and ${\cal L}_{t}^{+}$ and ${\cal L}_{t}^{++}$ become equal only when 
\begin{align}
\gamma_{ij,nl} = \frac{t_{il,n}v_{lj,n}}{\sum_{l'}t_{il',n}v_{l'\!j,n}}. \label{gammaeq}
\end{align}
From $\partial {\cal L}_{t}^{++}/\partial t_{il,n}\!=\!0$, we obtain 
\begin{align}
t_{il,n} = \left[ \frac{ \left( \frac{2}{\nu} + 1 \right) \sum_{j} \frac{1}{\alpha_{ij,n}}|y_{ij,n}|^{2}\gamma_{ij,nl}^{\frac{2}{p}+1}v_{lj,n}^{-\frac{2}{p}} }{ \sum_{j} \frac{1}{\beta_{ij,n}} v_{lj,n} } \right]^{\frac{p}{p+2}}. \label{tMU}
\end{align}
By substituting (\ref{betaeq}) and (\ref{gammaeq}) into (\ref{tMU}), we have the following update rule for $t_{il,n}$: 
\begin{align}
t_{il,n} \leftarrow t_{il,n} \left[ \frac{ \sum_{j} |y_{ij,n}|^2 \left( \frac{\nu}{\nu+2} \sigma_{ij,n}^{2} + \frac{2}{\nu+2} |y_{ij,n}|^{2} \right)^{-1} \sigma_{ij,n}^{-p} v_{lj,n} }{ \sum_{j} \sigma_{ij,n}^{-p}v_{lj,n} } \right]^{\frac{p}{p+2}}. \label{tMU2}
\end{align}
Similarly to (\ref{tMU2}), the update rule for $v_{lj,n}$ can be obtained as 
\begin{align}
v_{lj,n} \leftarrow v_{lj,n} \left[ \frac{ \sum_{i} |y_{ij,n}|^2 \left( \frac{\nu}{\nu+2} \sigma_{ij,n}^{2} + \frac{2}{\nu+2} |y_{ij,n}|^{2} \right)^{-1} \sigma_{ij,n}^{-p} t_{il,n} }{ \sum_{i} \sigma_{ij,n}^{-p}t_{il,n} } \right]^{\frac{p}{p+2}}. \label{vMU2}
\end{align}
These update rules are similar to those in $t$-NMF, but they include the new domain parameter $p$. 
After we update the parameters $t_{il,n}$ and $v_{lj,n}$, the model $\sigma_{ij,n}^{p}$ should be updated by (\ref{modelsigma}).

\SetAlgoNlRelativeSize{-1} 
\begin{algorithm}[t]
\footnotesize
Initialize $\bm{W}_{i}$ with identity matrix and $t_{il,n}$ and $v_{lj,n}$ with positive random values for all $i$, $l$, and $n$\tcp*{Initialization}
Calculate (\ref{separation}) and (\ref{modelsigma}) for all $i$, $j$, and $n$\tcp*{Update $y_{ij,n}$ and $\sigma_{ij,n}$}
\Repeat{converge}{
	Calculate (\ref{updateW1}) and (\ref{updateW2}) for all $i$ and $n$\tcp*{Update $\bm{w}_{i,n}$}
	Calculate (\ref{separation}) for all $i$, $j$, and $n$\tcp*{Update $y_{ij,n}$}
	Calculate $(\ref{tMU2})$ for all $i$, $l$, and $n$\tcp*{Update $t_{il,n}$}
	Calculate (\ref{modelsigma}) for all $i$, $j$, and $n$\tcp*{Update $\sigma_{ij,n}$}
	Calculate $(\ref{vMU2})$ for all $l$, $j$, and $n$\tcp*{Update $v_{lj,n}$}
	Calculate (\ref{modelsigma}) for all $i$, $j$, and $n$\tcp*{Update $\sigma_{ij,n}$}
	Calculate (\ref{ch3:norm_W})--(\ref{ch3:norm_t}) for all $i$, $j$, $l$, and $n$\tcp*{Normalization}
}
Calculate (\ref{PB}) for all $i$, $j$, and $n$\tcp*{Back-projection technique}
\caption{Algorithm for $t$-ILRMA}
\label{algorithmtILRMA}
\end{algorithm}

\begingroup
 \renewcommand{\arraystretch}{1}
\begin{table}[t]
\vspace{-5pt}
\caption{Music and speech sources obtained from SiSEC2011}
\vspace{7pt}
\label{usedSources}
\begin{center}
\scalebox{0.85}[0.85]{
\begin{tabular}{cccc}
\Hline
 \raisebox{-0.2ex}[0cm][0cm]{Signal}	& \raisebox{-0.2ex}[0cm][0cm]{Data name}												& \raisebox{-0.2ex}[0cm][0cm]{Source (1/2)}	\\ \hline
 \raisebox{-0.2ex}[0cm][0cm]{Music 1}		& \raisebox{-0.2ex}[0cm][0cm]{bearlin-roads}											& \raisebox{-0.2ex}[0cm][0cm]{acoustic\unl{~}guit\unl{~}main/vocals}	\\ \hline
 \raisebox{-0.2ex}[0cm][0cm]{Music 2}		& \raisebox{-0.2ex}[0cm][0cm]{another\unl{~}dreamer-the\unl{~}ones\unl{~}we\unl{~}love}	& \raisebox{-0.2ex}[0cm][0cm]{guitar/vocals}	\\ \hline
 \raisebox{-0.2ex}[0cm][0cm]{Music 3}		& \raisebox{-0.2ex}[0cm][0cm]{fort\unl{~}minor-remember\unl{~}the\unl{~}name}			& \raisebox{-0.2ex}[0cm][0cm]{violins\unl{~}synth/vocals}	\\ \hline
 \raisebox{-0.2ex}[0cm][0cm]{Music 4}		& \raisebox{-0.2ex}[0cm][0cm]{ultimate\unl{~}nz\unl{~}tour}								& \raisebox{-0.2ex}[0cm][0cm]{guitar/synth}	\\ \hline
 \raisebox{-0.2ex}[0cm][0cm]{Speech 1}	& \raisebox{-0.2ex}[0cm][0cm]{dev1\unl{~}female4}										& \raisebox{-0.2ex}[0cm][0cm]{src\unl{~}1/src\unl{~}2}	\\ \hline
 \raisebox{-0.2ex}[0cm][0cm]{Speech 2}	& \raisebox{-0.2ex}[0cm][0cm]{dev1\unl{~}female4}										& \raisebox{-0.2ex}[0cm][0cm]{src\unl{~}3/src\unl{~}4}	\\ \hline
 \raisebox{-0.2ex}[0cm][0cm]{Speech 3}	& \raisebox{-0.2ex}[0cm][0cm]{dev1\unl{~}male4}											& \raisebox{-0.2ex}[0cm][0cm]{src\unl{~}1/src\unl{~}2}	\\ \hline
 \raisebox{-0.2ex}[0cm][0cm]{Speech 4}	& \raisebox{-0.2ex}[0cm][0cm]{dev1\unl{~}male4}											& \raisebox{-0.2ex}[0cm][0cm]{src\unl{~}3/src\unl{~}4}	\\ 
\Hline
\end{tabular}
}
\vspace{-10pt}
\end{center}
\end{table}
\endgroup

\begin{figure}[t!]
\begin{center}
\vspace{3pt}
\includegraphics[width=6.7cm]{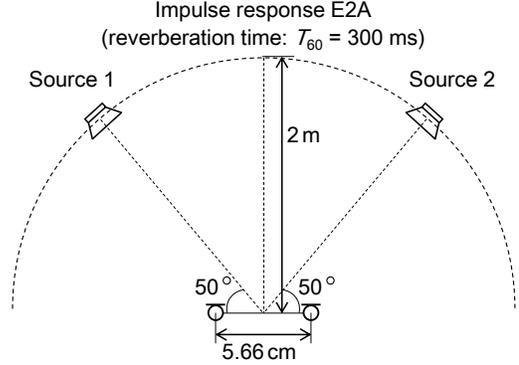}
\vspace{-12pt}
\caption{Recording conditions of impulse responses obtained from RWCP database.}
\label{RIR}
\end{center}
\vspace{-15pt}
\end{figure}

\begingroup
 \renewcommand{\arraystretch}{1}
\begin{table}[t]
\vspace{-1pt}
\caption{Relative computational times normalized by AuxIVA}
\vspace{3pt}
\label{relative}
\begin{center}
\scalebox{0.85}[0.85]{
\begin{tabular}{ccc}
\Hline
 \raisebox{-0.2ex}[0cm][0cm]{Method}				& \raisebox{-0.2ex}[0cm][0cm]{Two-source case}	& \raisebox{-0.2ex}[0cm][0cm]{Three-source case}	\\ \hline
 \raisebox{-0.2ex}[0cm][0cm]{AuxIVA~\cite{NOno2011_AuxIVA}}					& \raisebox{-0.2ex}[0cm][0cm]{1.00}				& \raisebox{-0.2ex}[0cm][0cm]{1.00}		\\ \hline
 \raisebox{-0.2ex}[0cm][0cm]{Proposed $t$-ILRMA}	& \raisebox{-0.2ex}[0cm][0cm]{1.46}				& \raisebox{-0.2ex}[0cm][0cm]{1.40}		\\ \hline
 \raisebox{-0.2ex}[0cm][0cm]{$t$-MNMF~\cite{KKitamura2016_tMNMF}}				& \raisebox{-0.2ex}[0cm][0cm]{8.83}				& \raisebox{-0.2ex}[0cm][0cm]{74.51}	\\ 
\Hline
\end{tabular}
}
\vspace{-10pt}
\end{center}
\end{table}
\endgroup

By iteratively calculating the update rules (\ref{updateW1}), (\ref{updateW2}), (\ref{tMU2}), and (\ref{vMU2}), the cost function 
(\ref{costt}) monotonically decreases, and the convergence is theoretically guaranteed. However, 
a scale ambiguity exists in the estimated signal $y_{ij,n}$ in ILRMA, and $\bm{W}_{i}$, $t_{il,n}$, and $v_{lj,n}$ should be normalized 
in each iteration as 
\begin{align}
\bm{w}_{i,n} \leftarrow&\ \bm{w}_{i,n}\eta_{n}^{-1}, \label{ch3:norm_W} \\
y_{ij,n} \leftarrow&\ y_{ij,n}\eta_{n}^{-1}, \label{ch3:norm_y} \\
\sigma_{ij,n}^{p} \leftarrow&\ \sigma_{ij,n}^{p} \eta_{n}^{-p}, \label{ch3:norm_r} \\
t_{il,n} \leftarrow&\ t_{il,n} \eta_{n}^{-p}, \label{ch3:norm_t}
\end{align}
where $\eta_{n}$ is an arbitrary sourcewise normalization coefficient, such as the sourcewise average power
\begin{align}
\eta_{n} = \sqrt{ \frac{1}{IJ}\sum_{i,j} |y_{ij,n}|^2 }. 
\end{align}
The signal scale of $y_{ij,n}$ can easily be restored by applying a 
back-projection technique after the cost function has converged, as 
\begin{align}
\hat{\bm{y}}_{ij,n} = \bm{W}_{i}^{-1} \left( \bm{e}_{n} \circ \bm{y}_{ij} \right), \label{PB}
\end{align}
where $\hat{\bm{y}}_{ij,n}$ is a scale-restored estimated source image and $\circ$ denotes element-wise multiplication. 
The algorithm for $t$-ILRMA is summarized in Algorithm~\ref{algorithmtILRMA}. 

\section{Experiments}
\label{exp}
\subsection{Conditions}
\label{exp:cond}


We confirmed the validity of the proposed generalization of ILRMA by conducting a BSS experiment using music and 
speech mixtures. The dry sources were obtained from SiSEC2011~\cite{SAraki2012_SiSEC} and are shown in Table~\ref{usedSources}. 
To simulate a reverberant mixture, the mixture signals were produced by convoluting the impulse response E2A ($T_{60}\!=\!300$~ms), 
which was obtained from the RWCP database~\cite{SNakamura2000_RWCP}, with each source. The recording conditions of the impulse responses 
are shown in Fig.~\ref{RIR}. The initial demixing matrix $\bm{W}_{i}$ was always set to the identity 
matrix, and the NMF parameters $t_{il,n}$ and $v_{lj,n}$ were initialized by random values. An STFT was performed using a 512-ms-long Hamming 
window with a 128 ms shift. The number of bases $L$ was set to five for music signals and two for speech signals, and the update rules were 
iterated 200 times. As the evaluation score, we used the signal-to-distortion ratio (SDR)~\cite{EVincent2006_BSSEval}, which 
indicates the overall separation quality. 

\subsection{Computational times compared with those of $t$-MNMF}
\label{exp:comptime}

\begin{figure}[t!]
\begin{center}
\vspace{3pt}
\includegraphics[width=8.4cm]{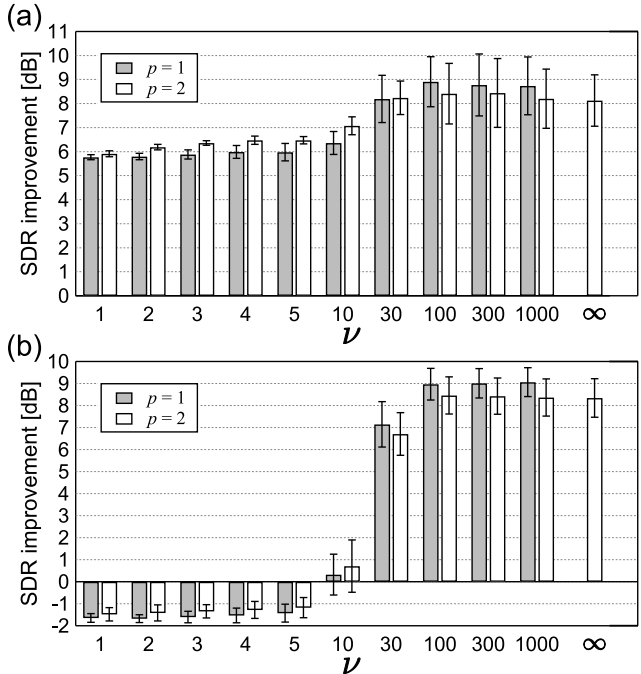}
\vspace{-12pt}
\caption{SDR improvements of conventional ILRMA and $t$-ILRMA: (a) Music 3 and (b) Speech 4.}
\label{results}
\end{center}
\vspace{-10pt}
\end{figure}

\begingroup
 \renewcommand{\arraystretch}{1}
\begin{table}[t]
\vspace{-5pt}
\caption{Average SDR improvements [dB] of $t$-ILRMA for music and speech signals}
\vspace{-7pt}
\label{resultAvg}
\begin{center}
\scalebox{0.85}[0.85]{
\begin{tabular}{ccccc}
\Hline
 \raisebox{-0.2ex}[0cm][0cm]{$\nu$}	& \raisebox{-0.2ex}[0cm][0cm]{Music ($p\!=\!1$)}	& \raisebox{-0.2ex}[0cm][0cm]{Music ($p\!=\!2$)}	& \raisebox{-0.2ex}[0cm][0cm]{Speech ($p\!=\!1$)}	& \raisebox{-0.2ex}[0cm][0cm]{Speech ($p\!=\!2$)} \\ \hline
 \raisebox{-0.2ex}[0cm][0cm]{1}					& \raisebox{-0.2ex}[0cm][0cm]{3.48}					& \raisebox{-0.2ex}[0cm][0cm]{3.32}				& \raisebox{-0.2ex}[0cm][0cm]{-0.18}				& \raisebox{-0.2ex}[0cm][0cm]{-0.15}	\\ \hline
 \raisebox{-0.2ex}[0cm][0cm]{2}					& \raisebox{-0.2ex}[0cm][0cm]{3.56}					& \raisebox{-0.2ex}[0cm][0cm]{3.44}				& \raisebox{-0.2ex}[0cm][0cm]{-0.24}				& \raisebox{-0.2ex}[0cm][0cm]{-0.22}	\\ \hline
 \raisebox{-0.2ex}[0cm][0cm]{3}					& \raisebox{-0.2ex}[0cm][0cm]{3.58}					& \raisebox{-0.2ex}[0cm][0cm]{3.50}				& \raisebox{-0.2ex}[0cm][0cm]{-0.26}				& \raisebox{-0.2ex}[0cm][0cm]{-0.22}	\\ \hline
 \raisebox{-0.2ex}[0cm][0cm]{4}					& \raisebox{-0.2ex}[0cm][0cm]{3.62}					& \raisebox{-0.2ex}[0cm][0cm]{3.61}				& \raisebox{-0.2ex}[0cm][0cm]{-0.29}				& \raisebox{-0.2ex}[0cm][0cm]{-0.30}	\\ \hline
 \raisebox{-0.2ex}[0cm][0cm]{5}					& \raisebox{-0.2ex}[0cm][0cm]{3.82}					& \raisebox{-0.2ex}[0cm][0cm]{3.99}				& \raisebox{-0.2ex}[0cm][0cm]{-0.30}				& \raisebox{-0.2ex}[0cm][0cm]{-0.30}	\\ \hline
 \raisebox{-0.2ex}[0cm][0cm]{10}				& \raisebox{-0.2ex}[0cm][0cm]{5.65}					& \raisebox{-0.2ex}[0cm][0cm]{6.16}				& \raisebox{-0.2ex}[0cm][0cm]{-0.13}				& \raisebox{-0.2ex}[0cm][0cm]{-0.19}	\\ \hline
 \raisebox{-0.2ex}[0cm][0cm]{30}				& \raisebox{-0.2ex}[0cm][0cm]{11.57}				& \raisebox{-0.2ex}[0cm][0cm]{11.02}			& \raisebox{-0.2ex}[0cm][0cm]{3.71}					& \raisebox{-0.2ex}[0cm][0cm]{3.44}		\\ \hline
 \raisebox{-0.2ex}[0cm][0cm]{100}				& \raisebox{-0.2ex}[0cm][0cm]{13.09}				& \raisebox{-0.2ex}[0cm][0cm]{12.66}			& \raisebox{-0.2ex}[0cm][0cm]{7.58}					& \raisebox{-0.2ex}[0cm][0cm]{6.87}		\\ \hline
 \raisebox{-0.2ex}[0cm][0cm]{300}				& \raisebox{-0.2ex}[0cm][0cm]{13.22}				& \raisebox{-0.2ex}[0cm][0cm]{12.78}			& \raisebox{-0.2ex}[0cm][0cm]{7.55}					& \raisebox{-0.2ex}[0cm][0cm]{7.18}		\\ \hline
 \raisebox{-0.2ex}[0cm][0cm]{1000}				& \raisebox{-0.2ex}[0cm][0cm]{13.27}				& \raisebox{-0.2ex}[0cm][0cm]{12.76}			& \raisebox{-0.2ex}[0cm][0cm]{7.74}					& \raisebox{-0.2ex}[0cm][0cm]{7.09}		\\ \hline
 \raisebox{-0.2ex}[0cm][0cm]{$\infty$}			& \raisebox{-0.2ex}[0cm][0cm]{-}					& \raisebox{-0.2ex}[0cm][0cm]{12.89}			& \raisebox{-0.2ex}[0cm][0cm]{-}					& \raisebox{-0.2ex}[0cm][0cm]{6.72}		\\ 
\Hline
\end{tabular}
}
\vspace{-10pt}
\end{center}
\end{table}
\endgroup

To clarify the advantage of ILRMA-based BSS in a determined situation ($N\!=\!M$), we compared the 
computational times of $t$-ILRMA and $t$-MNMF. 
The update calculation for the NMF parameters in each algorithm is almost the same, but the estimation of the spatial 
parameter ($\bm{W}_{i}$ for $t$-ILRMA and the spatial covariance for $t$-MNMF) is different. 
Although $t$-ILRMA requires one inverse of $\bm{W}_{i}\bm{U}_{i,n}$ for each $i$ and $n$, $t$-MNMF 
requires $J$ inverses and two eigenvalue decompositions of the $M\!\times\!M$ matrix. 
Table~\ref{relative} shows an example of relative computational times normalized by that of AuxIVA~\cite{NOno2011_AuxIVA}, 
where we used 
MATLAB 9.2 (64-bit) with an AMD Ryzen 7 1800X (8 cores and 3.6~GHz) 
CPU. From this table, we can confirm that the computational time of $t$-ILRMA 
does not increase significantly compared with that of IVA, whereas that of $t$-MNMF 
markedly increases.  
$t$-ILRMA was about six times faster than $t$-MNMF in the two-source case and about $53$ times faster in the 
three-source case. 

\begin{figure}[t!]
\begin{center}
\vspace{3pt}
\includegraphics[width=8.4cm]{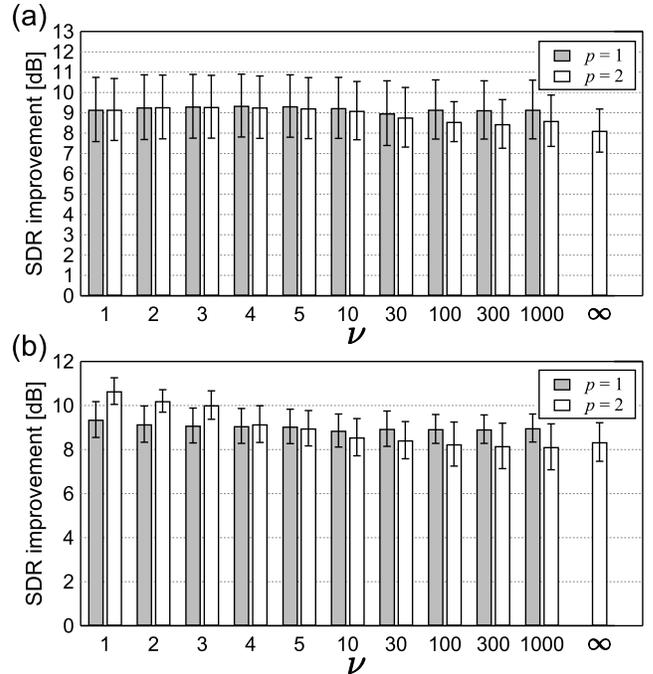}
\vspace{-12pt}
\caption{SDR improvements of conventional ILRMA and $t$-ILRMA with initialization: (a) Music 3 and (b) Speech 4.}
\label{resultsini}
\end{center}
\vspace{-10pt}
\end{figure}

\begingroup
 \renewcommand{\arraystretch}{1}
\begin{table}[t]
\vspace{-5pt}
\caption{Average SDR improvements [dB] of $t$-ILRMA with initialization for music and speech signals}
\vspace{-7pt}
\label{resultiniAvg}
\begin{center}
\scalebox{0.85}[0.85]{
\begin{tabular}{ccccc}
\Hline
 \raisebox{-0.2ex}[0cm][0cm]{$\nu$}		& \raisebox{-0.2ex}[0cm][0cm]{Music ($p\!=\!1$)}	& \raisebox{-0.2ex}[0cm][0cm]{Music ($p\!=\!2$)}	& \raisebox{-0.2ex}[0cm][0cm]{Speech ($p\!=\!1$)}	& \raisebox{-0.2ex}[0cm][0cm]{Speech ($p\!=\!2$)} \\ \hline
 \raisebox{-0.2ex}[0cm][0cm]{1}			& \raisebox{-0.2ex}[0cm][0cm]{13.15}				& \raisebox{-0.2ex}[0cm][0cm]{13.09}				& \raisebox{-0.2ex}[0cm][0cm]{6.93}					& \raisebox{-0.2ex}[0cm][0cm]{7.11}	\\ \hline
 \raisebox{-0.2ex}[0cm][0cm]{2}			& \raisebox{-0.2ex}[0cm][0cm]{13.31}				& \raisebox{-0.2ex}[0cm][0cm]{13.27}				& \raisebox{-0.2ex}[0cm][0cm]{6.97}					& \raisebox{-0.2ex}[0cm][0cm]{7.01}	\\ \hline
 \raisebox{-0.2ex}[0cm][0cm]{3}			& \raisebox{-0.2ex}[0cm][0cm]{13.34}				& \raisebox{-0.2ex}[0cm][0cm]{13.39}				& \raisebox{-0.2ex}[0cm][0cm]{6.89}					& \raisebox{-0.2ex}[0cm][0cm]{6.97}	\\ \hline
 \raisebox{-0.2ex}[0cm][0cm]{4}			& \raisebox{-0.2ex}[0cm][0cm]{13.27}				& \raisebox{-0.2ex}[0cm][0cm]{13.26}				& \raisebox{-0.2ex}[0cm][0cm]{6.99}					& \raisebox{-0.2ex}[0cm][0cm]{6.85}	\\ \hline
 \raisebox{-0.2ex}[0cm][0cm]{5}			& \raisebox{-0.2ex}[0cm][0cm]{13.25}				& \raisebox{-0.2ex}[0cm][0cm]{13.32}				& \raisebox{-0.2ex}[0cm][0cm]{6.94}					& \raisebox{-0.2ex}[0cm][0cm]{6.89}	\\ \hline
 \raisebox{-0.2ex}[0cm][0cm]{10}		& \raisebox{-0.2ex}[0cm][0cm]{13.41}				& \raisebox{-0.2ex}[0cm][0cm]{13.38}				& \raisebox{-0.2ex}[0cm][0cm]{7.04}					& \raisebox{-0.2ex}[0cm][0cm]{6.87}	\\ \hline
 \raisebox{-0.2ex}[0cm][0cm]{30}		& \raisebox{-0.2ex}[0cm][0cm]{13.39}				& \raisebox{-0.2ex}[0cm][0cm]{13.33}				& \raisebox{-0.2ex}[0cm][0cm]{7.25}					& \raisebox{-0.2ex}[0cm][0cm]{6.78}		\\ \hline
 \raisebox{-0.2ex}[0cm][0cm]{100}		& \raisebox{-0.2ex}[0cm][0cm]{13.39}				& \raisebox{-0.2ex}[0cm][0cm]{13.35}				& \raisebox{-0.2ex}[0cm][0cm]{7.29}					& \raisebox{-0.2ex}[0cm][0cm]{6.82}		\\ \hline
 \raisebox{-0.2ex}[0cm][0cm]{300}		& \raisebox{-0.2ex}[0cm][0cm]{13.39}				& \raisebox{-0.2ex}[0cm][0cm]{13.27}				& \raisebox{-0.2ex}[0cm][0cm]{7.21}					& \raisebox{-0.2ex}[0cm][0cm]{6.83}		\\ \hline
 \raisebox{-0.2ex}[0cm][0cm]{1000}		& \raisebox{-0.2ex}[0cm][0cm]{13.38}				& \raisebox{-0.2ex}[0cm][0cm]{13.32}				& \raisebox{-0.2ex}[0cm][0cm]{7.19}					& \raisebox{-0.2ex}[0cm][0cm]{6.84}		\\ \hline
 \raisebox{-0.2ex}[0cm][0cm]{$\infty$}	& \raisebox{-0.2ex}[0cm][0cm]{-}					& \raisebox{-0.2ex}[0cm][0cm]{12.89}				& \raisebox{-0.2ex}[0cm][0cm]{-}					& \raisebox{-0.2ex}[0cm][0cm]{6.72}		\\ 
\Hline
\end{tabular}
}
\vspace{-10pt}
\end{center}
\end{table}
\endgroup

\subsection{Results with random initialization}
\label{exp:resultsrand}

Fig.~\ref{results} shows an example of average SDR improvements and their standard deviations for various values of $\nu$, where 
the separation was performed 10 times with different random initializations for the parameters. Note that the result for 
$\nu\!\rightarrow\!\infty$ and $p\!=\!2$ corresponds to that for the conventional ILRMA assuming a complex Gaussian source generative model. 
From this result, we can confirm that the separation performance becomes stable and robust for the random initialization when 
$\nu$ is set to a small value. However, the performance is degraded for both music and speech signals when $\nu\!\leq\!10$. 
Only for the signals in Fig.~\ref{results} does the proposed method with $(\nu, p)\!=\!(100,1)$ for the music signal and $(\nu,p)\!=\!(1000,1)$ 
for the speech signal provide the best separation score. However, this tendency can vary with the 
dataset, namely, the optimal value of $\nu$ depends on the instrument or the speaker in the mixture signal. 
Table~\ref{resultAvg} shows the average scores of all music or speech signals. We can confirm that a higher value of $\nu$ and $p\!=\!1$ 
are always preferable for the separation. 

\subsection{Results with conventional ILRMA initialization}
\label{exp:resultsinit}

When $\nu$ is small, the Student's $t$-distribution approaches the Cauchy distribution, where the latter can ignore outlier components. In particular, 
Cauchy NMF is suitable for extracting significant bases from a truly low-rank data matrix contaminated by outlier noise~\cite{ALiutkus2015_CauchyNMF}. 
In the early stage of $t$-ILRMA iterations, the estimated spectrogram $|\bm{Y}_{n}|^{.p}$ includes almost all the source components because 
the initial demixing matrix is set to the identity matrix, and it is not a low-rank matrix even though each source spectrogram $|\bm{S}_{n}|^{.p}$ 
is truly low-rank. In such a case, $t$-ILRMA with a small value of $\nu$, such as Cauchy NMF, may not extract the useful bases for BSS, 
and the optimization will be trapped at a poor solution. 

To solve this problem, in this experiment, we apply conventional ILRMA ($t\!\rightarrow\!\infty$) in the early stage of iterations, 
then $t$-ILRMA with an arbitrary $\nu$ is applied in the late stage, where for $t$-ILRMA, the bases and activations are pretrained using 
the outputs of conventional ILRMA via $t$-NMF. Fig.~\ref{resultsini} and Table~\ref{resultiniAvg} show the average results of this approach, 
where conventional ILRMA is performed for the first 100 iterations and $t$-ILRMA is applied for the last 100 iterations. Compared with the 
previous results, a smaller value of $\nu$ tends to provide better results and to outperform conventional ILRMA, although 
the stability is not improved because of the conventional-ILRMA-based initialization. 

\section{Conclusion}
\label{conclusion}

In this paper, we generalized the source distribution assumed in ILRMA from a complex Gaussian distribution to a complex 
Student's $t$-distribution, which allows us to control the robustness to outlier components and includes the Cauchy distribution when $\nu\!=\!1$. 
The proposed $t$-ILRMA can outperform conventional ILRMA with an appropriate value of $\nu$. Also, initialization with a Gaussian 
assumption leads to further improvement for both music and speech BSS tasks.

\section{Acknowledgment}

This work was partly supported by ImPACT Program of Council for Science, SECOM Science and Technology Foundation, and JSPS KAKENHI Grant Number16H01735.


\end{document}